\documentclass[runningheads]{llncs}
\usepackage{graphicx}
\usepackage{hyperref}       
\usepackage{url}            
\usepackage{booktabs}       
\usepackage{amsfonts}       
\usepackage{nicefrac}       
\usepackage{xcolor}
\usepackage{microtype}      
\usepackage{mathtools} 
\usepackage{amsmath,mathptmx}
\usepackage[ruled, vlined]{algorithm2e}
\usepackage{subcaption}
\newtheorem{assumption}{Assumption}
\newtheorem{prop}{Proposition}

\DeclareMathOperator*{\E}{\mathbb{E}}

\DeclareMathOperator*{\R}{\mathbb{R}}

\usepackage{mathtools}
\graphicspath{ {./images/} }
\usepackage{dsfont}
\usepackage{}

\begin{document}

\title{On Poisoned Wardrop Equilibrium in Congestion Games }
\author{Yunian Pan\inst{1}\orcidID{0000-0002-7277-3657}
 \and Quanyan Zhu \inst{1}\orcidID{0000-0002-0008-2953}}
\authorrunning{Pan and Zhu}
\institute{New York University, Brooklyn, NY, USA; E-mail: \email{\{yp1170, qz494\}@nyu.edu}
}
\maketitle

\begin{abstract}
Recent years have witnessed a growing number of attack vectors against increasingly interconnected traffic networks.  Informational attacks have emerged as the prominent ones that aim to poison traffic data, misguide users, and manipulate traffic patterns. To study the impact of this class of attacks, we propose a game-theoretic framework where the attacker, as a Stackelberg leader, falsifies the traffic conditions to change the traffic pattern predicted by the Wardrop traffic equilibrium, achieved by the users, or the followers. 
The intended shift of the Wardrop equilibrium is a consequence of strategic informational poisoning. Leveraging game-theoretic and sensitivity analysis, we quantify the system-level impact of the attack by characterizing the concept of poisoned Price of Anarchy, which compares the poisoned Wardrop equilibrium and its non-poisoned system optimal counterpart. 
We use an evacuation case study to show that the Stackelberg equilibrium can be found through a two-time scale zeroth-order learning process and demonstrate the disruptive effects of informational poisoning, indicating a compelling need for defense policies to mitigate such security threats. 
  
 \end{abstract}

\section{Introduction}\label{intro}

With the rapid growth of the Internet-of-Things (IoT), there has been a significant number of vulnerable devices in the past decade, widening the cyber-physical attack surface of modern Intelligent Transportation Systems (ITS). 
For example, the adoption of IoT technologies for Vehicle-to-Vehicle (V2V), Vehicle-to-Infrastructure (V2I), and Infrastructure-to-Infrastructure (I2I) communications has enabled automated toll collection, traffic cameras and signals, road sensors, barriers, and Online Navigation Platforms (ONP) \cite{huq2017cyberattacks}. 
It, however, creates opportunities for attackers to disrupt the infrastructure by exploiting cyber vulnerabilities. 
A quintessential example of such attacks is the hijacking of traffic lights and smart signs. 
The recent work \cite{cerrudo2014hacking} demonstrates that due to lack of authentication, the wireless sensors and repeaters of the lighting control system can be accessed and manipulated through antenna, exposing serious vulnerabilities of the traffic infrastructure.


The impact of a local attack on the traffic systems propagates and creates a global disruption of the infrastructure. 
System-level modeling of cyber threats in traffic systems is crucial to understanding and assessing the consequences of cyber threats and the associated defense policies.  
One significant system-level impact is on the traffic conditions, including delays and disruptions. Attackers can launch a man-in-the-middle (MITM) on ONP systems to mislead the population to choose routes that are favored by the attackers.  
For instance, in 2014, two Israel students hacked the Google-owned Waze GPS app, causing the platform to report fake traffic conditions to its users; they used bot users to crowdsource false location information to the app, causing congestion \cite{popularnavihack2014}. 
A similar recent case happened in Berlin \cite{hackerjam2013}, where an artist loaded $99$ smartphones in the street, causing Google-Map to mark that street as having bad traffic.
It has been reported in \cite{google2020} that real-time traffic systems can be deceived by malicious attacks such as modified cookie replays and simulated delusional traffic flows. 

\begin{figure}[htbp]
    \centering
    \includegraphics[width = .7\textwidth]{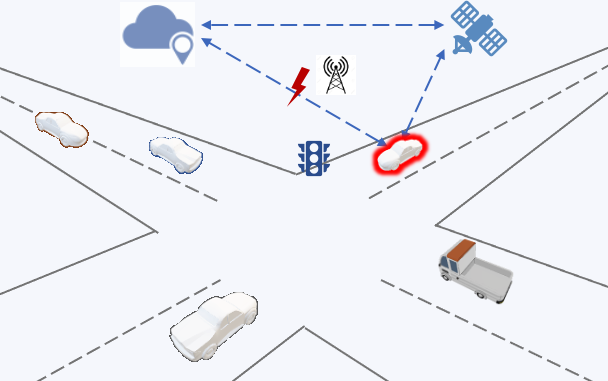}
    \caption{
    An example attack scenario: 
    a radio transmitter interferes the GPS communication channel, falsifying the user location information received by an ONP. ITS components, such as smart traffic signal and road cameras, can be hijacked to achieve the same goal.
    }
    \label{anexample}
\end{figure}

This class of attacks is referred to as {\it informational attacks} on traffic systems. They aim to exploit the vulnerabilities in the data and information infrastructures and strategically craft information to misguide users and achieve a target traffic condition.
The advent of information infrastructures and ONP has made user decisions more reliant on services offered by Google and Apple. 
This reliance has made the attack easily influence the populational behaviors in a much faster and more direct way. Fig. \ref{anexample} illustrates an example attack scenario.
The attack manipulates the information collected by an ONP, including traffic demand and travel latency, and misleads it to make false traffic prediction and path recommendations.

{\it Wardrop Equilibrium} (WE) \cite{wardrop1952road} has been widely used to predict the long-term behavioral patterns of the users and the equilibrium outcome of traffic conditions. 
 It is a natural system-level metric for the impact assessment of informational attacks.
Based on WE, we formulate a Stackelberg game  as our attack model. 
In this model, the attacker, or the leader, aims to disrupt the traffic system by poisoning the traffic conditions in a stealthy manner with bounded capabilities.
 To capture this strategic behavior, we let the attacker's utility consist of the cost of modifying the traffic conditions and the payoff of disruption outcome. In addition, stealthy information falsification attacks seek to satisfy flow conservation constraints to evade inconsistency check. 
 The best response of the users, or the followers, to such informational attacks is the path-routing equilibrium outcome subject to falsified traffic conditions, which are encapsulated by the poisoned traffic latency function and demand vector.
 We refer to the resulting behavioral pattern as the {\it Poisoned Wardrop Equilibrium} (PWE). 
 The disruptive effects of such attack is measured by the {\it Poisoned Price of Anarchy} (PPoA), which is the ratio of the aggregated latency under PWE to its non-poisoned system-optimal counterpart. 
 The local first-order stationary point is called differential Stackelberg equilibrium.

 The sensitivity analysis of the PWE and PPoA shows that the attacker's utility function is sufficiently smooth under regularity assumptions of the latency functions. We characterize the implicit relation between the PWE and the attack parameters, based on which we give an explicit expression for the gradient of attack utility. 
 By analyzing the attack gradient, we find that the existence of a differential Stackelberg equilibrium is determined by the weighting coefficient of attack payoff that captures the tradeoff between ``disruption'' and ``stealthiness''. 
We also uniformly characterize the locally Lipschitz parameters for both the attack utility and its gradient, which scale with a set of parameters, including the network size and topology, total traffic demand, and the smoothness level for the latency functions. 

We propose a zeroth order two-time scale learning algorithm to find the differential Stackelberg equilibrium and study the iterative adversarial behavior.
We approximate the attack gradient by sampling the aggregated latency outcome of PWE and give a polynomial
sample efficient guarantee for gradient approximation. 
We test our algorithm using an evacuation case study on a Sioux Falls network, where the attacker consistently learns to manipulate the information during the evacuation process through bandit feedback. 
We show that after several iterations, the PPoA of the entire traffic network converges to a PWE where the traffic flow concentrates on several particular edges, causing congestion and low road utilization rates. As
congestion games are ubiquitous not only in transportation networks but also in applications related to smart grid, distributed control, and wireless spectrum sharing, it is anticipated that similar attacks can occur in a broader range of scenarios, and there is a need for the development of secure and resilient mechanisms as future work. 

{\bf Content organization:} We briefly introduce  WE and some related works in Section \ref{rw}. In Section \ref{pf}, we present the model for WE and introduce two problem formulations corresponding to two fundamental principles, following which the attacker's problem is discussed. Section \ref{wediff} provides several theoretical aspects for attack objective function. In Section \ref{ag}, we explore the algorithmic development of the Stackelberg learning framework. We demonstrate the attack effects in Section \ref{sfdemo}.

\section{Related Work}\label{rw}

WE was introduced in 1952 \cite{wardrop1952road,beckmann1956studies} as an equilibrium model to predict the traffic patterns in transportation networks. 
The equilibrium concept is related to the notion of Nash equilibrium in game theory that was developed separately.  
Rosenthal in \cite{rosenthal1973class} introduced the class of congestion games and showed its existence of a pure-strategy Nash equilibrium.

There have been an extensive and growing literature that studies congestion games and their variants, and they have been used to model and understand the various technological impact on the transportation networks, including speed limits \cite{speedexogenous}, road pricing \cite{groot2014toward} or direct ONP assignment \cite{stackelbergweighted}. 
In these works, congestion games are subsumed as a building block to formulate Stackelberg games \cite{rockafellar2002} to design incentives, pricing, and policies. 
This work leverages the approach to create a formal framework to quantify and analyze the impact of the worst-case attack strategies on the transportation networks.

PoA is commonly used as a metric and analytical tool for congestion games. 
Cominetti et al. in \cite{DBLP:journals/corr/abs-1907-10101} have shown that PoA is a $C^1$ function of demand under certain conditions, which coincides to our results showing the smoothness of attack utility w.r.t. the demand poisoning parameter. 
Aligned with our discussions on the latency corruption, the effects of biased cost function have been investigated in \cite{meir2015playing}, their results are based on the notion of $(\lambda, \mu)$-smoothness \cite{roughgarden2010algorithmic}, which differs from our methods.
In general, PoA is sharply bounded by the condition number of the set of latency functions \cite{roughgarden2003price}, called the Pigou-bound. We refer the readers to \cite{Correa05onthe} for tighter analysis.
Specifically, for affine cost functions, this bound becomes $4/3$. Our numerical study shows that the inefficiency can be worse than the established results under informational attacks.

\section{Problem Formulation}\label{pf}

\subsection{Preliminary Background: Congestion Game and Wardrop Equilibrium}

Consider the traffic network as a directed graph $\mathcal{G} = (\mathcal{V}, \mathcal{E})$,
with the vertices $\mathcal{V}$ representing road junctions, and edges $\mathcal{E}$ representing road segments. 
We assume that $\mathcal{G}$ is finite, connected without buckles, i.e., the edges that connect a vertex to itself. The network contains the following elements:

\begin{itemize}
    \item $\mathcal{W} \subseteq \mathcal{V} \times \mathcal{V}$ is the set of distinct origin-destination (OD) pairs in the network; for $w \in \mathcal{W}$, $(o_w, d_w) \in \mathcal{V} \times \mathcal{V}$ is the OD pair; 
    \item $\mathcal{P}_w \subseteq \mathcal{P}(\mathcal{E})$ is the set of all directed paths from $o_w$ to $d_w$;
    \item $\mathcal{P} = \bigcup_{w\in\mathcal{W}}\mathcal{P}_w$ is the set of paths in a network, each $\mathcal{P}_w$ is disjoint;
    \item $Q \in \R^{|\mathcal{W}|}_{\geq 0}$ is the OD demand vector ,
    $Q_w$ represents the traffic demand between OD pair $w \in \mathcal{W}$;
    
    \item $q \in \R^{|\mathcal{E}|}_{\geq 0}$ is the edge flow vector, $q_e$ is the amount of traffic flow that goes through edge $e \in \mathcal{E}$
    \item $\mu \in \R^{|\mathcal{P}|}_+$ is the path flow vector, $\mu_p$ is the amount of traffic flow that goes through path $p \in \mathcal{P}$. 
    \item $\ell_e: \R_{\geq 0} \to \R_+ \ \ e \in \mathcal{E}$ is the cost/latency functions, determined by the edge flow. Let $\ell : \R^{|\mathcal{E}|}_{\geq 0} \to \R^{|\mathcal{E}|}_+$ denote the vector-valued latency function.
\end{itemize}

We assume that there is a set of infinite, infinitesimal players over this graph $\mathcal{G}$, denoted by a measurable space $(\mathcal{X}, \mathcal{M}, m)$. 
The players are non-atomic, i.e., $m(x) = 0 \ \ \forall x \in \mathcal{X}$; they are split into distinct populations indexed by the OD pairs, i.e., $\mathcal{X} = \bigcup_{w\in\mathcal{W}} \mathcal{X}_w$ and $\mathcal{X}_w \bigcap \mathcal{X}_{w^{\prime}} = \empty \ \ \forall w, w^{\prime} \in \mathcal{W}$.
For each player $x \in \mathcal{X}_w$, we assume that the path is fixed at the beginning, and thus the action of player $x$ is $A(x) \in \mathcal{P}_w$, which is $\mathcal{M}$-measurable. 
The action profile of all the players $\mathcal{X}$ induces the edge flows $q_e := \int_{\mathcal{X}} \mathds{1}_{\{e \in A(x)\}} m(dx) \ \ e \in \mathcal{E}$, and a path flow $\mu_p := {\int_{\mathcal{X}_w} \mathds{1}_{\{A(x) = p\}} m(dx)} \ \ p \in \mathcal{P}_w$, which are the fraction of players using edge $e$, and the fraction of players using $p \in \mathcal{P}_w$, respectively.
The path flow can also be interpreted as a mixed strategy played by a single centralized planner. 
By definition, a feasible flow pattern $(q, \mu) \in \R^{|\mathcal{E}|} \times \R^{|\mathcal{P}|}$ is constrained by \eqref{feasibleconstraint}:
\begin{equation}
    \label{feasibleconstraint}
    \begin{aligned}
       \Lambda \mu - Q & = 0  \\
     \Delta \mu - q & = 0 \\
       -\mu & \preceq 0 .
    \end{aligned}
\end{equation}
where $\Lambda \in \R^{|\mathcal{W}| \times |\mathcal{P}|}$, $\Delta \in \R^{|\mathcal{E}| \times |\mathcal{P}|}$ are the path-demand incidence matrix and the path-edge incidence matrix, respectively, which are defined in \eqref{pathdemandedge}. 
The two matrices only depend on the topology of network $\mathcal{G}$.  
\begin{equation}
    \label{pathdemandedge}
    \Lambda_{w p}=\left\{\begin{array}{ll}1 & \text { if } p \in \mathcal{P}_{w} \\ 0 & \text { otherwise }\end{array} \quad \text { and } \quad \Delta_{e p}=\left\{\begin{array}{ll}1 & \text { if } e \in p \\ 0 & \text { otherwise }\end{array}\right.\right. .
\end{equation}

The utility function for a single player is the aggregated cost for the path she selects, $\ell_p (\mu) = \sum_{e \in p} \ell_e(q_e)$. Note that the path latency is a function of the path flow vector $\mu$.
We hereby impose the first assumption about the edge latency functions.
\begin{assumption}[($\ell$-Regularity)]
\label{latencyassumption}
 For all $e \in \mathcal{E}$, the latency functions $\ell_e$ are $l_0$-Lipschitz continuous, twice differentiable with $\ell_e^{\prime} (q_e) > 0 $, and $\ell_e^{\prime\prime}(q_e) \geq 0$ for $q_e \geq 0$. In addition, $\ell^{\prime}_e$ are $l_1$-Lipschitz continuous and $\ell^{\prime\prime}_e$ is bounded by $l_1$. 
\end{assumption}
The path latency $\ell_p$ can be bounded by $D(\mathcal{G})c_0$, where $D(\mathcal{G})$ the diameter of the graph $\mathcal{G}$, and $c_0 : = \| \ell\|_{\infty} = \max_{e \in \mathcal{E}} \ell_e (D)$.
This congestion game $\mathcal{G}_c$ is thus encapsulated by the triplet $(\mathcal{X}, \ell, \mathcal{P})$. 

\subsection{System Optimum and Wardrop Equilibrium} 

In the seminar work \cite{wardrop1952road}, Wardrop proposed two different principles, leading to two solution concepts.

\begin{itemize}
    \item \textit{Wardrop's first principle (Nash equilibrium principle)}: Players aim to minimize their own travel cost, i.e., for a mixed strategy $\mu$ to be a Nash equilibrium, whenever a path $ \mu_p > 0$ is chosen for the OD pair $w$, it holds that $\ell_p(\mu) \leq \ell_{p'}(\mu)$ $\forall p' \in \mathcal{P}_w$, implying that every flow has the same latency.
    
    \item \textit{Wardrop's second principle (social optimality principle)}: Players pick routes cooperatively such that the overall latency is minimized. The coordinated behaviors minimize the aggregated system performance $\sum_{e \in \mathcal{E}} q_e \ell_e(q_e)$ under proper constraints.
\end{itemize}

We hereby formalize the notion of { \it System Optimum} (SO) and WE. 
Definition \ref{wardropsecond} follows Wardrop's second principle, characterizing the cooperative behaviors of individuals that minimize the aggregated latency.
 
\begin{definition}[System Optimum (SO)]\label{wardropsecond}
 The socially optimal routing $(q^{\star}, \mu^{\star})$ is a feasible flow pattern that optimizes the social welfare by minimizing the aggregated latency $S(q) = \sum_{e \in \mathcal{E}} q_e \ell_e(q_e)$, obtained from the optimization problem \eqref{socialopt}
 \begin{equation} \label{socialopt}
     \begin{aligned}
             \min_{q, \mu } \quad &\sum_{e \in \mathcal{E}} q_e \ell_e(q_e) \\
     \text{s.t.} \quad &  (q, \mu) \in F_Q
     \end{aligned}
 \end{equation}
 where $F_Q := \{(q, \mu) \in \R^{|\mathcal{E}|} \times \R^{|\mathcal{P}|} |  (q, \mu) \text{ satisfies \eqref{feasibleconstraint}.} \}$
\end{definition}

By assumption \ref{latencyassumption}, problem \eqref{socialopt} is strictly convex in $\R^{|\mathcal{E}|}$, admitting a strict global minimum edge flow $q^{\star}$, the corresponding path flow set $\pmb{\mu}^{\star}$ is generally the non-unique solution to the linear equation $\Delta \mu = q^{\star}$, satisfying \eqref{feasibleconstraint}. 
The optimal aggregated latency is denoted by $S^{\star} := S(q^{\star})$.

The Nash equilibrium, on the other hand, exploits the self-interest nature of the individuals in a transportation network. 
Definition \ref{wardropfirst} follows Wardrop's first principle, characterizing the non-cooperative behaviors of individuals that minimize their own latency. 

\begin{definition}[Wardrop Equilibrium (WE)] \label{wardropfirst}
A flow pattern $(q, \mu)$ is said to be a Wardrop Equilibrium (WE), if it satisfies $(q, \mu) \in F_Q $, and for all $w \in \mathcal{W}$:
 \begin{itemize}
     \item $\ell_p (\mu) = \ell_{p'}(\mu)$ for all $p, p' \in \mathcal{P}_w$ with $ \mu_p, \mu_{p'} >0$;
     \item $\ell_p(\mu) \geq \ell_p(\mu)$ for all $p, p' \in \mathcal{P}_w$ with $ \mu_p >0$ and $\mu_{p'} = 0$.
 \end{itemize}
 Equivalently, WE can be characterized as the minimizer of the following convex program: 
\begin{equation} \label{beckmanopt}
\begin{aligned}
      \min_{ q, \mu } \quad & \sum_{e \in \mathcal{E}} \int_{0}^{q_e} \ell_e(z) dz\\
    \text{s.t.} \quad & (q, \mu ) \in F_Q,
\end{aligned}
\end{equation}
where $\sum_{e \in \mathcal{E}} \int_{0}^{q_e} \ell_e(z) dz = : J(q)$ is called the Beckman potential.
\end{definition}


Since, by assumption \ref{latencyassumption}, $\ell_e$ is strictly increasing, the equilibrium edge flow $q^*$ is uniquely defined; the corresponding equilibrium path flow set $\pmb{\mu}^*$ is generally the non-unique solution to the linear equation $\Delta \mu = q^*$, satisfying \eqref{feasibleconstraint}.

\subsection{Stackelberg Congestion Security Game}

This section formulates a Stackelberg congestion security game. We consider
an attacker who manipulates latency and demand data to mislead the ONP and its users. 
%
To capture this malicious behavior, we introduce a pair of attack parameters $ (\theta, d) \in (\Theta \times \mathcal{D})$ as the attack action, which parameterize two global traffic condition operators, $\Phi_{\theta}: \Theta \times \R^{|\mathcal{E}|} \times \R^{|\mathcal{P}|} \to \R^{|\mathcal{E}|} \times \R^{|\mathcal{P}|} $ and $\Phi_d: \mathcal{D} \times \R^{|\mathcal{W}|} \to \R^{| \mathcal{W}|}$. 
The flow operator $\Phi_{\theta}$ modifies the real-time traffic flow to poison the latency function; 
the demand operator $\Phi_d$ poisons the traffic demand prediction.

After the poisoning, the demand prediction and latency function are corrupted to be $\tilde{Q} : = \Phi_d \cdot Q$ and $\tilde{\ell} = \ell \circ \Phi_{\theta} $, respectively.
We hereby introduce the {\it $(\theta,d)$-Poisoned Wardrop Equilibrium} ($(\theta,d)$-PWE) as described in \ref{pwe}.


\begin{definition}[$(\theta,d)$-PWE]\label{pwe}
 A flow pattern $(q, \mu)$ is said to be a $(\theta,d)$-PWE, if it is a solution to the problem \eqref{beckmanopt}, with the latency function being $\tilde{\ell} = \Phi_{\theta} \circ \ell$ and the OD demand vector being $\tilde{Q} = \Phi_d \cdot Q$. 
 The equilibrium edge flow and path flow set are denoted by $q^*(\theta, d)$, and $ \pmb{\mu}^*(\theta, d)$, respectively. 
 
\end{definition}

\begin{figure}[htbp]
    \centering
    \includegraphics[width=.9\textwidth]{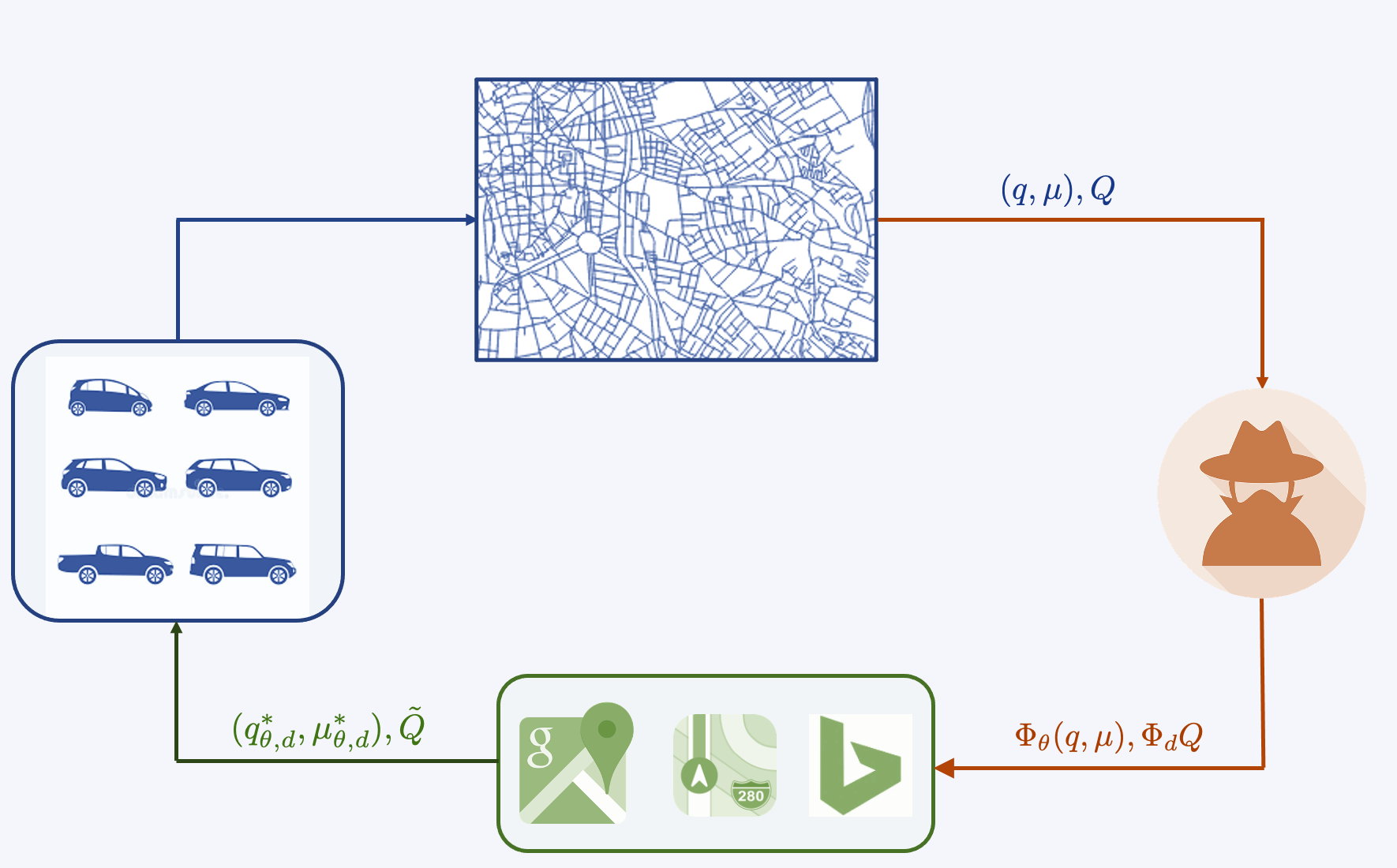}
    \caption{An illustration of the $(\theta,d)$-PWE-formation loop: the attacker stealthily intercepts the communication channel that collects traffic conditions, forcing the formation of traffic flow that is favored by the attacker.}
    \label{attackfeedback}
\end{figure}

As illustrated in Fig. \ref{attackfeedback}, the corruption of real-time traffic conditions, the poisoned path recommendation by ONP, and the user path selection (the formation of PWE) form a closed-loop system that is interfered by the attacker. 

To quantify the disruption caused by such informational attack, we introduce the notion of $(\theta,d)$-{\it Poisoned Price of Anarchy} ($(\theta,d)$-PPoA). 
\begin{definition}\label{ppoadef}
The ratio of aggregated latency at $(\theta,d)$-PWE 
 to the aggregated latency at non-poisoned SO is called  $(\theta,d)$-PPoA , i.e.: 
 \begin{equation}\label{ppoa}
      \text{$(\theta,d)$-PPoA} = \frac{\sum_{e \in \mathcal{E}} q^*_e(\theta,d) \ell_e(q^*_e(\theta,d))}{\sum_{e \in \mathcal{E}} q^{\star}_e \ell_e (q^{\star}_e)}.
 \end{equation}
\end{definition}

Now, we are ready to define attacker's cost function and complete the attack model.
We give two formulations in the sequel, based on the malicious manipulation of edge flow and path flow, respectively.

\subsubsection{Edge Flow Poisoning}

In this case, the attacker corrupts the latency function through a global edge flow operator $\Phi_{\theta}:  \Theta \times \R^{|\mathcal{E}|} \mapsto \R^{|\mathcal{E}|}$. For simplicity, we consider the attack operators to be matrices of proper dimensions, i.e.,  $\Phi_{\theta} \in \R^{|\mathcal{E}| \times |\mathcal{E}|}$, and $ \Phi_d \in \R^{|\mathcal{W}| \times |\mathcal{W}|}$.

The operators $\Phi_{\theta}$ and $\Phi_d$ have the following interpretation. 
Through data manipulation, the fraction $\Phi_{\theta; i, j}$ of traffic flow in edge $i$ is redistributed to edge $j$; the fraction $\Phi_{d, i, j}$  of demand between OD pair $i$ is redirected to OD pair $j$.
It is reasonable to let $\|\Phi_{\theta}\|_{op}$ and $\|\Phi_d\|_{op}$ be $1$ such that the flow and demand corruption cannot be identified by checking the norm of the flow and demand vectors. The set of column-stochastic matrix satisfies such a constraint.

The problem \eqref{edgeflowattackopt} is to optimize the attack utility $\mathcal{L}: \Theta \times \mathcal{Q} \times \R^{|\mathcal{E}|}   \mapsto \R$. 
The utility $\mathcal{L}$ contains two terms. 
The attack cost term is measured by the $\|\cdot\|_F$  norm of deviation from ``no-attack'' to ``attack''; 
the attack payoff term is the $(\theta,d)$-PPoA weighted by parameter $\gamma$, which measures the disruption of the transportation network. 

\begin{equation} \label{edgeflowattackopt}
    \begin{aligned}
          \min_{ (\theta, d) \in \Theta \times \mathcal{Q} , \  q = q^*(\theta, d)} \ \  &\mathcal{L}\left((\theta, d) , q\right) : =  \frac{1}{2} (\| \Phi_{\theta} - I \|^2_F + \| \Phi_d - I\|^2_F) - \gamma \frac{\sum_{e \in \mathcal{E}} q_e \ell_e(q_e)}{\sum_{e \in \mathcal{E}} q^{\star}_e \ell_e (q^{\star}_e)} \\
          \text{s.t. }  & \quad  \Phi_{\theta}^{\top} \mathds{1} = \mathds{1} ,
           \quad  \\
           & \quad  \Phi_{\theta; i,j} \geq 0 \quad  \forall i, j \in 1, \ldots, |\mathcal{E}|, \\
           & \quad \Phi_d^{\top} \mathds{1} = \mathds{1} , \\
            &\quad  \Phi_{d; i,j} \geq 0 \quad \forall  i, j \in 1, \ldots, |\mathcal{W}|.
     \end{aligned}
\end{equation}

\subsubsection{Path Flow Poisoning}

 In this case, the attacker corrupts the latency function through a global path flow operator $\Phi_{\theta}: \Theta \times \R^{|\mathcal{P}|} \mapsto \R^{|\mathcal{P}|}$. 
 Let $\Phi_{\theta} \in \R^{|\mathcal{P}|\times |\mathcal{P}|}$ and $\Phi_d \in \R^{|\mathcal{W}| \times |\mathcal{W}|}$, with similar path flow and demand operating interpretation. 
 Writing the $(\theta,d)$-PPoA term with respect to the path flow, we can restate the problem as in \eqref{pathflowattackopt}:
\begin{equation} \label{pathflowattackopt}
    \begin{aligned}
          \min_{ (\theta, d) \in \Theta \times \mathcal{Q} } \sup_{ \mu \in \pmb{\mu}^*(\theta, d)} \ \  &\mathcal{L}\left((\theta, d) , \mu \right) : =  \frac{1}{2} (\| \Phi_{\theta} - I \|^2_F + \| \Phi_d - I\|^2_F) - \gamma \frac{\sum_{p \in \mathcal{P}} \mu_p \ell_p(\mu) }{\sum_{p \in \mathcal{P}} \mu^{\star}_p \ell_p (\mu^{\star})}\\
          \text{s.t. }  & \quad  \Phi_{\theta}^{\top} \mathds{1} = \mathds{1} ,
           \quad  \\
           & \quad  \Phi_{\theta; i,j} \geq 0 \quad  \forall i, j \in 1, \ldots, |\mathcal{P}|, \\
           & \quad \Phi_d^{\top} \mathds{1} = \mathds{1}, \\
            &\quad  \Phi_{d; i,j} \geq 0 \quad \forall i, j \in 1, \ldots, |\mathcal{W}| ,
     \end{aligned}
\end{equation}
where we take the supremum over the path flow set of $(\theta,d)$-PWE. 
One can verify that the normalizing denominator $\sum_{p \in \mathcal{P}} \mu^{\star}_p \ell_p (\mu^{\star}) = \sum_{e \in \mathcal{E}} q^{\star}_e\ell_e(q^{\star}_e) $, i.e., while the one edge flow may correspond to multiple path flows, the aggregated latency remains the same.

Since in general the optimal path flow $\pmb{\mu}^*(\theta,d)$ is a set-valued mapping, we focus on problem \eqref{edgeflowattackopt} for analytical convenience in the sequel. 
For the Stackelberg game defined in \eqref{edgeflowattackopt}, we refer to the constraint set as $\mathcal{C}$. 
The convexity of the mathematical program \eqref{edgeflowattackopt} can not be determined due to the implicity of parameterization $(\theta,d)$-PWE. Assuming that the parameterization yields sufficient smoothness conditions, we adopt the first-order local stationary point as the solution concept, called {\it Differential Stackelberg Equilibrium} (DSE), as described in Definition \ref{dse}. 

\begin{definition}[Differential Stackelberg Equilibrium (DSE)] \label{dse}
A pair $\left((\theta^*,d^*), (q, \pmb{\mu})\right)$ with $(\theta^*, d^*)\in\mathcal{C}$, $(q,\pmb{\mu}) = (q^*(\theta^*, d^*), \pmb{\mu}^*(\theta^*,d^*))$ being the $(\theta^*,d^*)$-PWE, is said to be a Differential Stackelberg Equilibrium (DSE) for the Stackelberg game defined in \eqref{edgeflowattackopt}, if 
$\nabla_{\theta,d} \mathcal{L} = 0$, and $\nabla^2_{\theta,d} \mathcal{L}$ is positive definite.  
\end{definition}

In practice, we consider the explicit case where $\Phi_{\theta}$ is a matrix in $\R^{|\mathcal{E}|\times |\mathcal{E}|}$ (or $\R^{|\mathcal{P}|\times |\mathcal{P}|}$) and is parameterized by $\theta \in \Theta =  \R^{|\mathcal{E}|^2 }$ (or $\R^{|\mathcal{P}|^2 }$) such that $vec(\Phi_{\theta}) = \theta$, and $\Phi_d$ is parameterized by $d \in \mathcal{D} = \R^{|\mathcal{W}|^2 }$ such that $vec(\Phi_d) = d$. In this case, $\mathcal{C}$ is a compact and convex set. Later on, we use the operator $\operatorname{Proj}_{\mathcal{C}}( \theta, d)$ to represent the Euclidean projection onto $\mathcal{C}$, i.e., $\operatorname{Proj}_{\mathcal{C}}( \theta, d) = \arg \min_{z \in \mathcal{C}} \| z - ( \theta, d)\|^2$.

\section{Sensitivity Analysis} \label{wediff}

\subsection{Smoothness of $(\theta, d)$-PWE}

Let $\Theta, \mathcal{D}$ be open sets, for some fixed parameter $(\theta, d) \in \Theta \times \mathcal{D}$, a unique minimizer $q^*(\theta, d)$ of the parameterized Beckman program \eqref{parabeck} is uniquely determined.
\begin{equation}\label{parabeck}
\begin{aligned}
        \min_{q, \mu} \ \  &  J( (q, \mu) |\theta, d) := \sum_{e \in \mathcal{E}} \int_{0}^{q}(\ell \circ \Phi_{\theta})_e(z) d\\
   \text{s.t.}  \quad & (\Phi_{\theta} q, \mu ) \in F_{\Phi_d Q}. 
\end{aligned}
\end{equation}
 
To study the sensitivity of $\mathcal{L}$ and $q^{\star}(\theta,d)$ to the perturbations of $\theta$ and $d$, we reduce the feasibility set for the parameterized version of Beckman program \eqref{beckmanopt} to the $q$ variable first. 
In doing so, we give Lemma \ref{charfeasibility}.
\begin{lemma}\label{charfeasibility}
~Given attack parameter $\theta, d$, define the feasible set of edge flow 
\begin{equation*}
    \pmb{q}_{\theta, d} := \{ q \in \mathbb{R}^{|\mathcal{E}|} \  \big\vert \  \exists \mu \text{ such that } (\Phi_{\theta}q ,\mu) \in F_{\Phi_d Q} \},
\end{equation*} which has the following properties:
 \begin{itemize}
     \item[(a)] There exists $A \in \R^{r \times |\mathcal{E}|}$ and $B \in \R^{r \times |\mathcal{W}|}$ of proper dimensions, with $r$ depending only on $ \mathcal{G}$, such that
     \begin{equation*}
          \pmb{q}_{\theta,d} = \{ q \in \mathbb{R}^{| \mathcal{E}|} \ \big\vert \ A \Phi_{\theta}q \leq B\Phi_d Q\} .
     \end{equation*}
     \item[(b)] Any $q \in \pmb{q}_{\theta,d}$ is bounded by 
     \begin{equation*}
         \| q\| \leq D\sqrt{|\mathcal{E}|}.
     \end{equation*}
     \item[(c)] There exists a constant $l_d$ such that for any $d^{\prime}, d \in \mathcal{D}$ and $ q \in \pmb{q}_{\theta,d}$ there exists $q^{\prime} \in \pmb{q}_{\theta, d^{\prime}}$ satisfying
     \begin{equation*}
         \|q^{\prime} - q \| \leq l_d \|d^{\prime} - d \|.
     \end{equation*}
 \end{itemize}
\end{lemma}
By lemma \ref{charfeasibility}, the feasibility set $\pmb{q}_{\theta,d}$ can be projected onto $q$-space as a linear inequality constraint on $q$-variable, which is bounded and local Lipschitz smooth w.r.t. $d$.

\begin{lemma}\label{continuity}
  Let $q^*(\theta, d)$ be the unique minimizer of \eqref{parabeck}. Then, at each $(\bar{\theta}, \bar{d}) \in \Theta \times \mathcal{D}$, there exists $\varepsilon$ such that for all $(\theta, d) \in B_{\varepsilon}(\bar{\theta}, \bar{d})$:
  \begin{itemize}
      \item[(a)] The edge flow at $(\theta,d)$-PWE, $q^*(\theta, d)$ is continuous, i.e., for any sequence $ (\theta_n, d_n) \to (\bar{\theta}, \bar{d}), n \in \mathbb{N}$, we have $q^*(\theta_n, d_n) \to q^*(\theta, d)$. In addition, there exists a Lipschitz constant $ l_q > 0$ that is related to $\|B \Phi_d \|$ such that
      \begin{equation}
          \label{lipqthetad}
          \|q^*(\theta, d) - q^*(\bar{\theta}, \bar{d})\| \leq l_q \|(\theta,d ) - (\bar{\theta}, \bar{d}) \| 
      \end{equation}
      
      \item[(b)] The poisoned aggregated latency function $ S(q^*(\theta,d))$ is locally Lipschitz continuous, i.e., 
      \begin{equation*}
        \|  S(q^*(\theta, d)) - S(q^*(\bar{\theta}, \bar{d})) \| \leq  (c_0 + l_0 D) l_q \sqrt{|\mathcal{E}|}\| (\theta, d) - (\bar{\theta}, \bar{d})\| .
      \end{equation*}
  \end{itemize}
\end{lemma}
The Lipschitz constant in \ref{continuity} (b) has the following interpretation.
The smoothness level of the poisoned aggregated latency function scales with three factors:  the upper estimate scale of latency ($\|\ell\|_{\infty} \text{ and } l_0 D$), the network size ($\sqrt{|\mathcal{E}|}$), and the smoothness level of $(\theta,d)$-PWE ($l_q$). 
This Lipschitz constant directly implies the smoothness level of $(\theta,d)$-PPoA.

\subsection{Differentiability of $(\theta, d)$-PWE}
By lemma \ref{charfeasibility}, the feasibility set can be reduced to a linear inequality constraint. 
Define the $(\theta, d)$-poisoned Lagrangian:
\begin{equation}\label{plagrangian}
    L(q, \lambda, \theta, d) = \sum_{e\in \mathcal{E}} \int_{0}^{(\Phi_{\theta} q)_e} \ell_e (z) dz + \lambda^{\top}(A \Phi_{\theta}q - B\Phi_d Q).
\end{equation}
The KKT condition states that a vector $\tilde{q} \in \R^{|\mathcal{E}|}$ is the solution $q^*(\theta,d)$ if and only if there exists $\tilde{\lambda} \in \R^{r}$ such that:
\begin{equation*}
    \begin{aligned}
          A \Phi_{\theta}\tilde{q} - B\Phi_d Q & \preceq 0 \\
           \tilde{\lambda}_i & \geq 0, \quad i = 1, \ldots, r  \\
           \tilde{\lambda}_i (A \Phi_{\theta} \tilde{q} - B \Phi_d Q)_i & = 0,  \quad i = 1, \ldots, r \\
         \sum_{e^{\prime} \in \mathcal{E}}  \Phi_{\theta; e, e^{\prime}}^{\top} \ell_{e^{\prime}}( (\Phi_{\theta} \tilde{q} )_{e^{\prime}}) + (\Phi_{\theta}^{\top } A^{\top} \tilde{\lambda} )_e & = 0 , \quad e = 1, \ldots,  |\mathcal{E}| ,
    \end{aligned}
\end{equation*}
To apply Implicit Function Theorem (IFT) to the poisoned Beckman program \eqref{beckmanopt}, we define the vector-valued function $g = \nabla_{(q, \lambda)} L$,
\begin{equation}\label{ifth}
    g( \tilde{q}, \tilde{\lambda}, \theta, d) = \begin{bmatrix}
\sum_{e^{\prime} \in \mathcal{E}}  \Phi_{\theta; e^{\prime}, 1} \ell_{e^{\prime}}( (\Phi_{\theta} \tilde{q} )_{e^{\prime}}) + (\Phi_{\theta}^{\top } A^{\top} \tilde{\lambda} )_1 \\ 
\ldots\\
 \sum_{e^{\prime} \in \mathcal{E}}  \Phi_{\theta; e^{\prime}, |\mathcal{E}|} \ell_{e^{\prime}}( (\Phi_{\theta} \tilde{q} )_{e^{\prime}}) + (\Phi_{\theta}^{\top } A^{\top} \tilde{\lambda} )_{|\mathcal{E}|}\\
\operatorname{diag}(\lambda) (A \Phi_{\theta} \tilde{q} - B\Phi_d Q) \end{bmatrix},
\end{equation}
where $\operatorname{diag}(\cdot)$ transforms the vector $\lambda$ into the matrix with $\lambda_i$ being the diagonal entries. For a candidate WE solution $(\tilde{q}, \tilde{\lambda})$ such that $g(\tilde{q}, \tilde{\lambda}, \theta, d ) = 0$, we define the partial Jacobian w.r.t. variable $(q, \lambda)$:
\begin{equation}\label{dqlambda}
    \mathrm{D}_{(q, \lambda)} g(\tilde{q}, \tilde{\lambda}, \theta, d)=\begin{bmatrix}
\mathrm{D}_{q} \nabla_{q} L(\tilde{q}, \tilde{\lambda}, \theta, d) &  \Phi_{\theta}^{\top} A^{\top}  \\
\operatorname{diag}(\tilde{\lambda}) A \Phi_{\theta} & \operatorname{diag}(A \Phi_{\theta} \tilde{q} - B\Phi_d Q) 
\end{bmatrix} ,
\end{equation}
where the first diagonal term $$
\mathrm{D}_{q} \nabla_{q} L(\tilde{q}, \tilde{\lambda}, \theta, d) = [\sum_{e^{\prime} \in [\mathcal{E}]}  \Phi_{\theta; e^{\prime}, i}  \Phi_{\theta; e^{\prime}, j}  \ell^{\prime}_{e^{\prime}}( (\Phi_{\theta} \tilde{q} )_{e^{\prime}})]_{i, j \in [\mathcal{E}]} = \Phi_{\theta}^{\top} \nabla_q \tilde{\ell},$$ is positive definite according to assumption \ref{latencyassumption}. By Shur's complement, one can verify that if $\{ i \ \big \vert  \tilde{\lambda}_i = 0 \text{ and } (A \Phi_{\theta} \tilde{q} - \Phi_d Q)_i = 0\} =\emptyset$, the partial Jacobian is non-singular.

The partial Jacobian w.r.t. variable $(\theta, d)$ is
\begin{equation}\label{jacobianthetad}
    \mathrm{D}_{(\theta, d)} g(\tilde{q}, \tilde{\lambda},  \theta, d)
    =\begin{bmatrix} 
    \mathrm{D}_{\theta} \nabla_{q} L(\tilde{q}, \tilde{\lambda},  \theta, d)  & \mathrm{D}_{d} \nabla_{q} L(\tilde{q}, \tilde{\lambda},  \theta, d)\\ 
    \operatorname{diag}(\tilde{\lambda}) \mathrm{D}_{\theta}  (A \Phi_{\theta} \tilde{q}) & - \operatorname{diag}(\tilde{\lambda}) \mathrm{D}_d(B\Phi_dQ) 
    \end{bmatrix}.
\end{equation}
 
Lemma \ref{iftbeckman} gives the local differentiability result for $(\theta,d)$-PWE.

\begin{lemma}[IFT for $(\theta,d)$-PWE] \label{iftbeckman}
 Let $g(\tilde{q}, \tilde{\lambda}, \theta, d) = 0$,  if the set $\{ i \ \big \vert \tilde{\lambda}_i = 0 \text{ and } (A \Phi_{\theta} \tilde{q} - \Phi_d Q)_i = 0\} =\emptyset$, then $\mathrm{D}_{(q, \lambda)} g(\tilde{q}, \tilde{\lambda},  \theta, d )$ is non-singular, then the solution mapping for WE \eqref{beckmanopt} has a single-value localization $q^*(\theta, d)$ around $(\tilde{q}, \tilde{\lambda})$, which is continuously differentiable in the neighbor of $(\theta,d)$ with partial Jacobian satisfying: 
 \begin{equation}
 \mathrm{D}_{\theta} q^*(\theta, d)=-\mathrm{D}_{(q, \lambda)} g(\tilde{q}, \tilde{\lambda},  \theta, d)^{-1} \mathrm{D}_{\theta} g(\tilde{q}, \tilde{\lambda}, \theta, d) \quad\quad \forall \theta \in \Theta,
\label{qwrttheta}
 \end{equation}
 and 
  \begin{equation}
 \mathrm{D}_{d} q^*(\theta, d)=-\mathrm{D}_{(q, \lambda)} g(\tilde{q}, \tilde{\lambda},  \theta , d)^{-1} \mathrm{D}_{d} g(\tilde{q}, \tilde{\lambda}, \theta, d) \quad\quad \forall d \in \mathcal{D} ,
\label{qwrtd}
 \end{equation}
 where $\mathrm{D}_{(q, \lambda)} g(\tilde{q}, \tilde{\lambda},  \theta, d)$ is defined in \eqref{dqlambda}, and $[\mathrm{D}_{\theta} g(\tilde{q}, \tilde{\lambda}, \theta, d), \mathrm{D}_{d} g(\tilde{q}, \tilde{\lambda}, \theta, d)]$ is defined in \eqref{jacobianthetad}.
 
\end{lemma}
A similar derivation for the path flow case is given in Appendix \ref{appendixa}.

\subsection{Characterizing Attacker Objective}

Equipped with Lemma \ref{iftbeckman}, we arrive at the explicit expression for $\nabla \mathcal{L}$ in Theorem \ref{attackgrad}. 
\begin{theorem}\label{attackgrad}
 For problem \eqref{edgeflowattackopt}, the gradient of $\mathcal{L}$ w.r.t. $\theta$ is: 
 \begin{equation}\label{attackedgegradtheta}
    \nabla_{\theta}  \mathcal{L} = \theta - vec(I_{|\mathcal{E}|})  - \frac{\gamma}{S^{\star}} \sum_{e \in \mathcal{E}} \left( q_e^*(\theta, d) \frac{d \ell_e(z)}{d z}\big\vert_{ q_e^*(\theta,d)} +  \ell_e(q_e^*(\theta,d))  \right) \nabla_{\theta}q_e^*(\theta,d),
 \end{equation}
 where $\nabla_{\theta}q_e^*(\theta,d)$ is the transpose of $\mathrm{D}_{\theta}q_e^*(\theta,d)$ defined in \eqref{qwrttheta}.
 
 The gradient of $\mathcal{L}$ w.r.t. $d$ is:
 \begin{equation}\label{attackedgegradd}
    \nabla_{d}  \mathcal{L} = d - vec(I_{|\mathcal{W}|})  - \frac{\gamma}{S^{\star}} \sum_{e \in \mathcal{E}} \left( q_e^*(\theta, d) \frac{d \ell_e(z)}{d z}\big\vert_{q_e^*(\theta,d)} +  \ell_e(q_e^*(\theta,d))  \right) \nabla_{d}q_e^*(\theta,d),
 \end{equation}
 where $\nabla_{d}q_e^*(\theta,d)$ is the transpose of $\mathrm{D}_{d}q_e^*(\theta,d)$ defined in \eqref{qwrtd}.
\end{theorem}

Theorem \ref{attackgrad} also indicates that the existence of a DSE can be controlled by the weighting factor $\gamma$. 
To see this, we first notice that the first-order condition $\nabla \mathcal{L}$ may not be achievable within $\mathcal{C}$ when $\gamma$ is too large.
For the second-order condition, observe that the Hessian $\nabla^2_{\theta}\mathcal{L}$ takes the form similar to an $M$-matrix, i.e., $\nabla^2_{\theta} \mathcal{L} = I - \gamma H$, where $H$ is:
\begin{equation*}
    M =  \frac{1}{S^{\star}}\nabla_{\theta}  \sum_{e \in \mathcal{E}} \left( q_e^*(\theta, d) \frac{d \ell_e(z)}{d z}\big\vert_{ q_e^*(\theta,d)} +  \ell_e(q_e^*(\theta,d))  \right) \mathrm{D}_{\theta}q_e^*(\theta,d).
\end{equation*}
Under proper scaling of $\gamma$, the positive definiteness of $\nabla^2_{\theta}\mathcal{L}$ can be guaranteed, given the spectral radius of $M$ is strictly less than $\frac{1}{\gamma}$ everywhere in $\Theta \times \mathcal{D}$. The same analysis can be applied to $\nabla^2_d \mathcal{L}$.

The weighting parameter $\gamma$ also plays a role in balancing the local sensitivities of attack cost and payoff, as described in Theorem \ref{attackobjlip}.

\begin{theorem}\label{attackobjlip}
 The attacker objective function $\mathcal{L}$ is $L_0$-locally Lipshcitz continuous w.r.t. its argument $\theta$ and $d$, where $L_0$ is:
 \begin{equation}\label{lipconst}
     L_0 =  (\sqrt{2}   +  \gamma \frac{(c_0 + l_0 D) l_q}{S^{\star}} )\sqrt{|\mathcal{E}|} .
 \end{equation}
\end{theorem}

$L_0$ consists of two terms: 
one is the smoothness level of quadratic cost that scales with the network size factor $\sqrt{|\mathcal{E}|}$; 
one is the smoothness level of $(\theta, d)$-PPoA that scales with not only $\sqrt{|\mathcal{E}|}$, but also the ratio between Lipschitz constants of $S(q^*(\theta,d))$ and $S^{\star}$. 
It can be computed that $S^{\star}$ roughly scales with $c_0 D \sqrt{|\mathcal{E}|}$, 
hence $\gamma l_q$ must scale with $\sqrt{|\mathcal{E}|}$ to match the sensitivities of attack cost and payoff. 
This again indicates that the selection of weighting factor $\gamma$ is non-trivial.

The gradient smoothness is an important condition for the convergence analysis of gradient-based algorithms. 
Determining the Lipschitz constant of $\nabla \mathcal{L}$ requires the upper estimates of $\|\nabla_{\theta, d} q_e^{*}(\theta,d)\|_{op}$, which in turn requires the lower eigenvalue estimates $ \lambda_{\min} (\mathrm{D}_{(q, \lambda)} g)$ and upper eigenvalue estimates $\lambda_{\max}(\mathrm{D}_{\theta,d} g)$.
Intuitively, the boundedness of the partial Jacobians of $g$ can be guaranteed by the regularity assumption of $\ell$ and $\Phi_{\theta}$, which is already made in our context. 
We end this section with Lemma \ref{gradientlip}, which characterize the gradient smoothness under the regularity assumptions of $\|\nabla_{\theta, d} q_e^{*}(\theta,d)\|_{op}$.

\begin{lemma}\label{gradientlip}
 Given $\|\nabla_{\theta, d} q_e^{*}(\theta,d)\|_{op}$ is bounded by $C_0$ and  $C_1$-locally Lipschitz continuous, the attacker objective gradient $\nabla_{\theta} \mathcal{L}$ is $L_1$-locally Lipschitz continuous w.r.t. its argument $\theta$, where $L_1$ is:
 \begin{equation}
     L_1 = 1 + \frac{\gamma}{S^{\star}}\left(C_0 lq (l_0 + \ell^{\prime}(D))+ C_1 c_0 + D \sqrt{|\mathcal{E}|} (C_0 l_1 l_q + C_1 \ell^{\prime}(D) )\right)\sqrt{|\mathcal{E}|} .
 \end{equation}
\end{lemma}

\section{Algorithmic Development} \label{ag}

\subsection{Consistent Attack as a Stackelberg Learning Process}

Projected gradient-based method is a standard approach to find a first-order stationary point or a DSE. 
As shown in algorithm \ref{firstorderlearn}, the two-time scale Stackelberg learning procedure requires the attacker to have access to the first-order oracle, which gives the zeroth and first-order information of the edge latencies, the traffic flow at PWE, and the partial Jacobians of $g$. 
\begin{algorithm}[htbp]
\SetKwInOut{Input}{Input}
\SetAlgoLined
\Input{Admissible initial parameter $\theta, d$, learning rate $\eta$;} 
\While{not done}{ 
  Attacker initiates attack $\Phi_{\theta}, \Phi_d$\;
  \While{Attacking}{
     Players form $(\theta,d)$-PWE according to $\ell\circ \Phi_{\theta}$ and demand $\Phi_d Q$\;
     Attacker observes $(q^*, \mu^*)(\theta,d)$ and performs projected gradient updates\;
     \begin{equation}
     \begin{aligned}
                \theta \leftarrow   \mathrm{Proj}_{\mathcal{C}}[\theta - \eta \nabla_{\theta} \mathcal{L}] \quad
                d \leftarrow \mathrm{Proj}_{\mathcal{C}}[d - \eta \nabla_{d} \mathcal{L}] 
     \end{aligned}
     \end{equation}
    }
}
\caption{First-Order Poisoning}
\label{firstorderlearn}
\end{algorithm}

This framework can be viewed as a closed-loop feedback learning process. Every attack iteration is a period of $(\theta,d)$-PWE formation, given the poisoning configuration as input; the first-order oracle reveals the result for the attacker to consistently adjust the poisoning strategy.

The first-order oracle is oftentimes unavailable in practice. WHAT QUESTION IS HERE? The question is whether the attacker is able to approximately find the Stackelberg differential equilibria through bandit-feedback, i.e., the aggregated latency results of $(\theta,d)$-PWE.
To this end, we define two smoothed versions of attacker utility,  
\begin{equation}
\begin{aligned}
        \mathcal{L}^{\theta}_r((\theta,d), q^* ) = \E_{ u \sim  \mathbb{B}^{\theta}_r} [\mathcal{L}((\theta + u, d), q^*) ], \\ \mathcal{L}^{d}_r((\theta,d), q^* ) = \E_{ v \sim \mathbb{B}^d_r} [\mathcal{L}((\theta , d + v), q^*) ], 
\end{aligned}
\end{equation}
where $u,v$ are uniformly sampled from $r$-radius Frobenius norm balls $\mathbb{B}^{\theta}_r, \mathbb{B}^{d}_r$ with proper dimensions.
As smoothed functions, $\mathcal{L}^{\theta}_r, \mathcal{L}^d_r$ have Lipschitz constants no worse than $L$ for all $r > 0$, and their gradients, by standard volume argument from \cite{flaxman2004online} Lemma 2.1, 
\begin{equation}\label{zerograd}
    \begin{aligned}
          \nabla_{\theta}  \mathcal{L}^{\theta}_r((\theta,d), q^* ) = \frac{dim(\Theta)}{r^2}\E_{ u \sim  \mathbb{S}^{\theta}_r} [\mathcal{L}((\theta + u, d), q^*) u],  \\
          \nabla_{d}  \mathcal{L}^{d}_r((\theta,d), q^* ) = \frac{dim(\mathcal{D})}{r^2}\E_{ v \sim  \mathbb{S}^{\theta}_r} [\mathcal{L}((\theta , d + v), q^*) v], 
    \end{aligned}
\end{equation}
where $\mathbb{S}^{\theta}_r, \mathbb{S}^{d}_r$ are $r$-radius spheres of proper dimensions. 

Equipped with the smoothness results and \eqref{zerograd},  by standard concentration inequalities, we show that it suffices to use polynomial numbers of samples to approximate the gradients.

\begin{prop}[Gradient Approximation Efficiency] \label{gradefficiency}
Given a small $\epsilon > 0$, one can find fixed polynomials $h_r(1/ \epsilon)$, $h_{sample}(dim(\Theta), 1/ \epsilon)$, $h_{sample}(dim(\mathcal{D}), 1/ \epsilon)$, for $r \leq h_r(1/\epsilon)$, with $m \geq \max \{h_{sample}(dim(\Theta), 1/ \epsilon), h_{sample}(dim(\mathcal{D}), 1/ \epsilon) \}$ samples of $u_1$, $\ldots$, $u_m$ and  $v_1, \ldots, v_m$, with probability at least $1 - (d/\epsilon)^{-d}$ the  sample averages
\begin{equation} \label{sampleavg}
\frac{dim(\Theta)}{mr^2}\sum_{i=1}^m\mathcal{L}((\theta + u_i,d), q^*) u_i , \quad \frac{dim(\mathcal{D})}{mr^2}\sum_{i=1}^m\mathcal{L}((\theta,d + v_i), q^*) v_i
\end{equation}
are $\epsilon$ close to $\nabla_{\theta}\mathcal{L}$ and $\nabla_d \mathcal{L}$, respectively.
\end{prop}

Leveraging the one-point gradient approximation technique, we propose the derivative-free Algorithm \ref{zeroorderlearn} as an alternative to Algorithm \ref{firstorderlearn}. 
This algorithm asynchronously perturbs the parameters $\theta$ and $d$ to obtain the one-point gradient estimates.

\begin{algorithm}[htbp]
\SetKwInOut{Input}{Input}
\SetAlgoLined
\Input{Admissible initial parameter $\theta, d$, learning rate $\eta$, sample size $m$, radius $r$;} 
\While{not done}{ 
  Attacker initiates attack $\Phi_{\theta}, \Phi_d$\;
  \While{Attacking}{
   \For{$i = 1,\ldots, m$}{
     Sample $(\theta,d)$-PWE for searching directions $u_i \sim \mathbb{S}^{\theta}_r$, $v_i \sim \mathbb{S}^d_r$, obtain: 
     \begin{equation*}
    \mathcal{L}^{\theta}_i = \mathcal{L} (\operatorname{Proj}_{\mathcal{C}}(\theta + u_i, d ), q^*) , \quad \mathcal{L}^{d}_i = \mathcal{L} (\operatorname{Proj}_{\mathcal{C}}(\theta, d + v_i), q^*).
     \end{equation*}
     }
    Projected gradient updates:
     \begin{equation}
                \theta \leftarrow   \mathrm{Proj}_{\mathcal{C}}[\theta - \eta \frac{dim(\Theta)}{mr^2}\sum_{i=1}^m \mathcal{L}_{\theta}^i u_i] \quad 
                d \leftarrow \mathrm{Proj}_{\mathcal{C}}[d - \eta  \frac{dim(\mathcal{D})}{mr^2}\sum_{i=1}^m \mathcal{L}_d^i v_i] .
     \end{equation}
    }
}
\caption{Zeroth-Order Poisoning}
\label{zeroorderlearn}
\end{algorithm}

By projecting perturbed $\theta, d$ to the constraint set $\mathcal{C}$, we ensure the feasibility of the perturbed attack strategies when sampling $\mathcal{L}^{\theta}_i$ and $\mathcal{L}^d_i$. 

Algorithm \ref{zeroorderlearn} can proceed without the aid of first-order oracle, but it requires the number of samples to be polynomial w.r.t. the smoothness level of attack utility.

\section{Case Study} \label{sfdemo}

\begin{figure}[htbp]
\centering
\begin{subfigure}{.5\textwidth}
\centering
  \includegraphics[width=\textwidth]{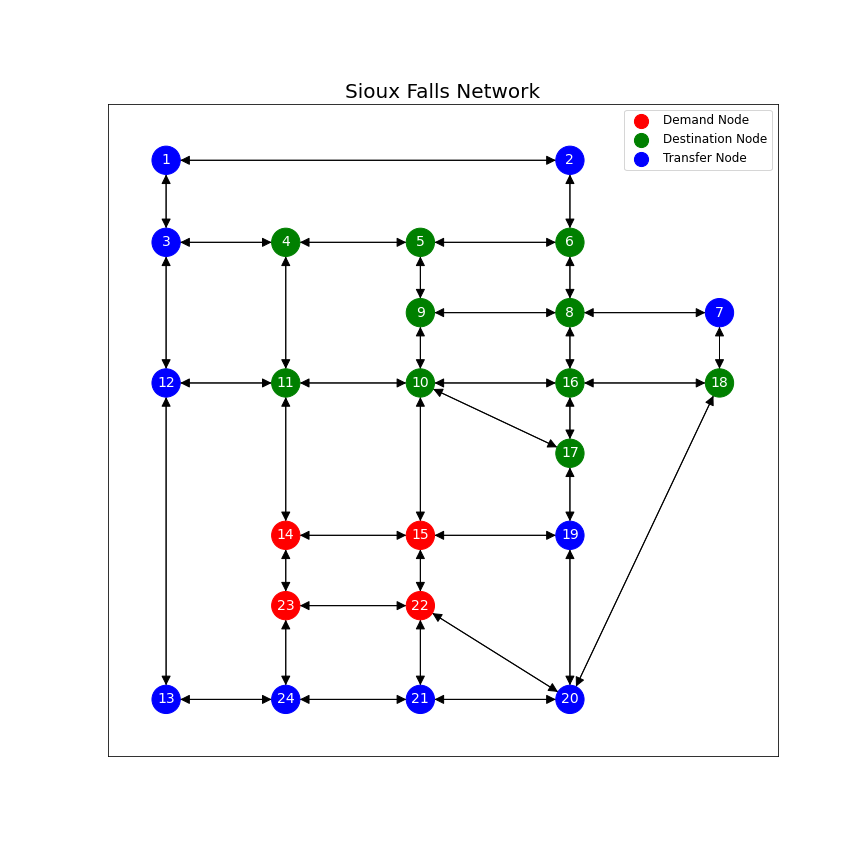}
  \label{sfsnetworktopo}
\end{subfigure}%
\begin{subfigure}{.5\textwidth}
\centering
  \includegraphics[width=.8\textwidth]{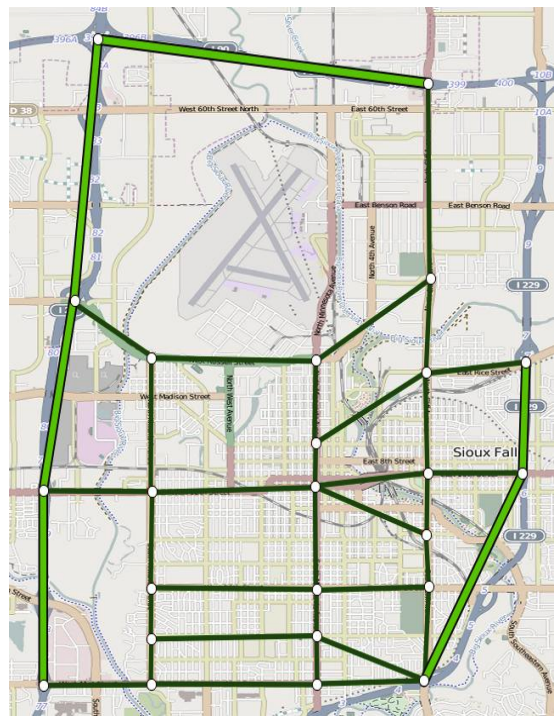}
\end{subfigure}
\caption{The topological (left) and geographical layout \cite{chakirov2014enriched} (right) of the Sioux Falls city, South Dakota transportation network. 
 The red nodes represent the locations where people need to be evacuated, the green nodes represent the evacuation shelters, and the blue nodes represent transfer locations. 
}
\label{sf}
\end{figure}

Through an emergency evacuation case study \cite{ng2010hybrid}, over the classical Sioux Falls, South Dakota Transportation Network \cite{siouxfalls} (Fig. \ref{sf} (b)), we test our Stackelberg learning algorithm and demonstrate the attack effects.

In our example, the evacuation lasts for one month. 
During each day, a total of 34200 individuals are transported from emergency locations (the red nodes) (14), (15), (22), and (23), to shelter places (the green nodes) (4), (5), (6), (8), (9), (10), (11), (16), (17), and (18). 
The transportation network data, including node attributes, free travel time, and road capacity, etc., are available at  \cite{github}.  
The edge latency is given by the standard Bureau of Public Roads (BPR) function:
\begin{equation}\label{bprfunc}
    \ell_{e}\left(q_{e}\right)=t_{e}^{f}\left(1+\alpha\left(\frac{q_{e}}{\mathrm{C}_{e}}\right)^{\beta}\right),
\end{equation}
where $t_e^f$ is the free time for edge $e$, $\mathrm{C}_e$ is the road capacity for edge $e$, and $\alpha, \beta$ are some parameters.

\begin{figure}
    \centering
    \includegraphics[width=.65\textwidth]{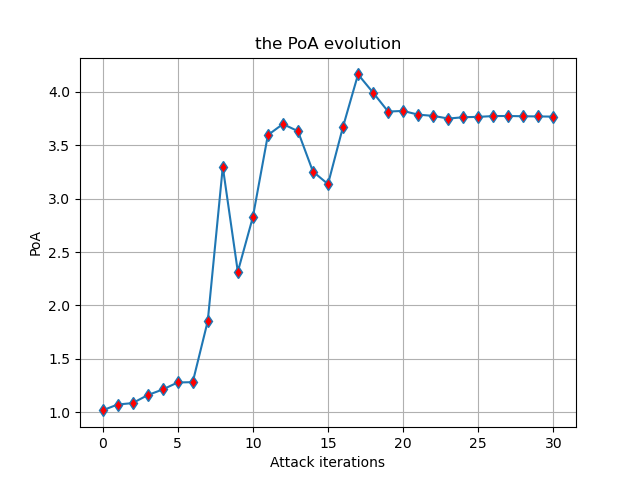}
    \caption{The evolution of PPoA: after $15$ days, the PPoA of this attack scenario reaches above $4$; and the process is stabilized at day-$20$ and attains the PPoA around $3.6$.}
    \label{evopoa}
\end{figure}

The attacker's goal is to slow down the evacuation process through latency and demand poisoning.  
The attacker can launch multiple attacks during one day, for each attack, the aggregated latency at the corresponding PWE is revealed as an observation to the attacker.
These observations are then used to update the attack strategy.
The weighting factor $\gamma$ and sample size $m$ are both picked to scale with $\sqrt{|\mathcal{E}|}$, where $|\mathcal{E}| = 76$ is the total edge number. 
An annealing factor of $0.95$ is used for the learning rate. 
We sample perturbations $u_i$ and $v_i$ from a standard normal distribution for practical purposes.  
The PPoA evolution curve is shown in Fig. \ref{evopoa}. 

Fig. \ref{evopoa}  shows that a stealthy attacker can decrease the efficiency of WE, pushing it far away from the SO. 
The convergence of the Stackelberg learning process implies the finding of a DSE.

\begin{figure}
\centering
\begin{subfigure}[b]{\textwidth}
   \includegraphics[width=1\linewidth]{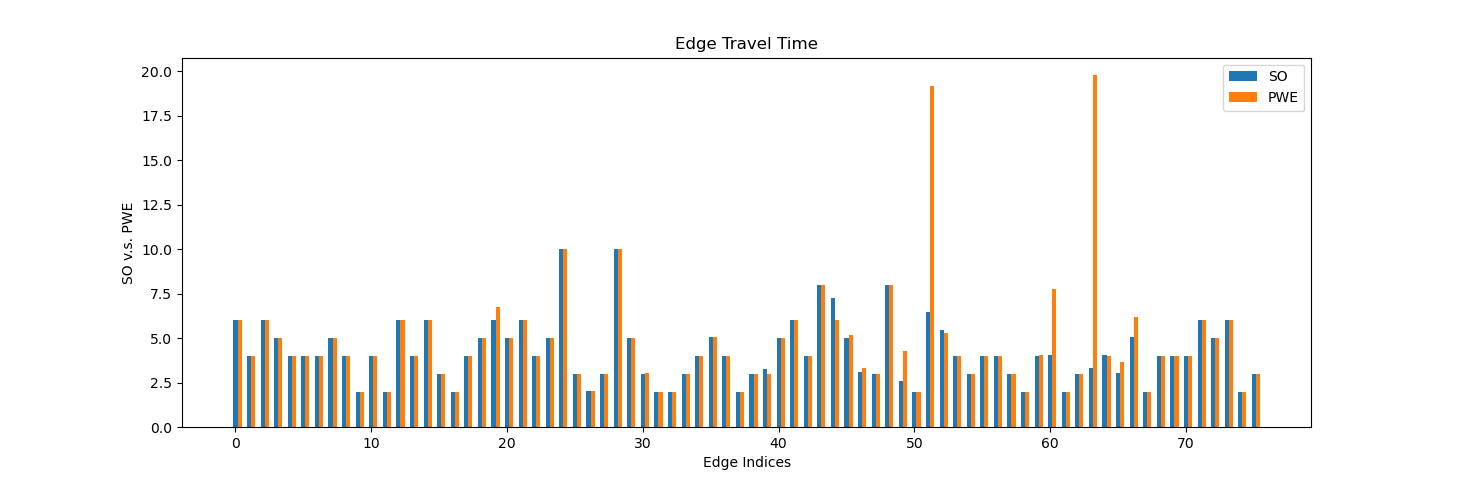}
   \caption{This bar chart compares the edge travel time caused by SO and PWE of the last day. 
   Edges indexed by 49, 51, 60, and 63 are experiencing significant traffic delays.}
   \label{edgetime} 
\end{subfigure}

\begin{subfigure}[b]{\textwidth}
   \includegraphics[width=1\linewidth]{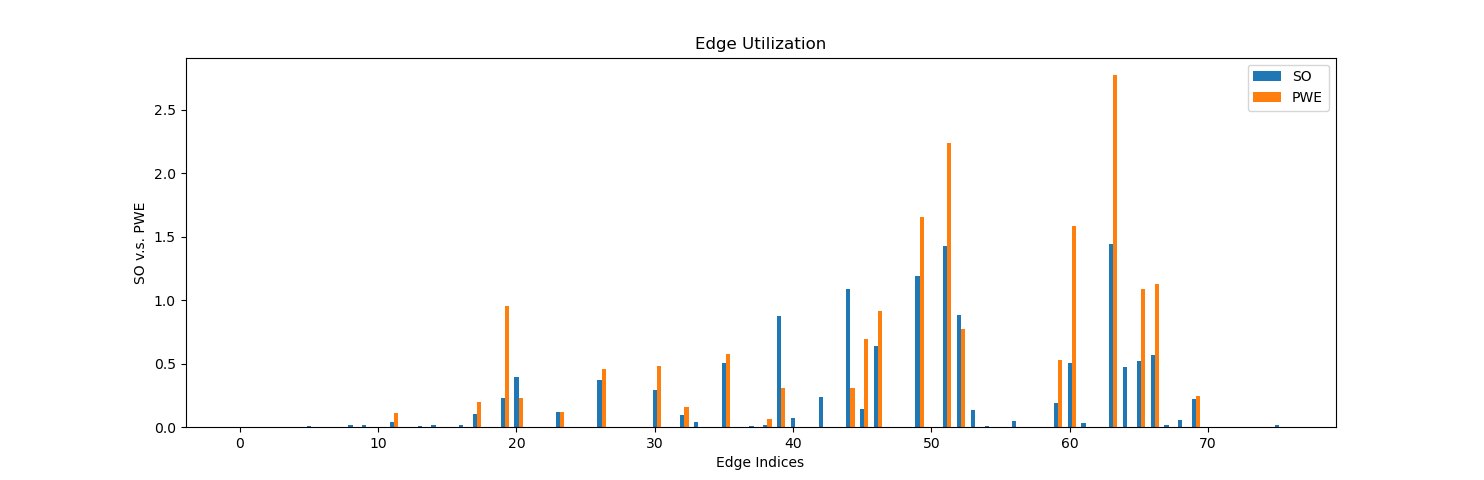}
   \caption{This bar chart compares the edge utilization rate caused by SO and PWE of the last day. 
   Corresponding to Fig. \ref{edgetime}, traffic flows in edges indexed by 49, 51, 60, and 63  are significantly larger than the edge capacities.}
   \label{edgeutilization}
\end{subfigure}
\caption[Two edge efficiency plots]{Two edge efficiency comparison bar charts comparing the edge time and edge utilization rate caused by SO and PWE of the last attack.}
\label{edgeefficiency}
\end{figure}

We compare the edge efficiencies of SO and PWE in Fig. \ref{edgeefficiency}. Fig. \ref{edgetime} shows the comparison of latency function values for each edge, given by \eqref{bprfunc} and the assigned edge flow; Fig. \ref{edgeutilization} shows the comparison of utilization ratio between the actual flow on that edge and its road capacity, $q_e / \mathrm{C}_e$. 

Fig. \ref{edgetime} shows that at the end of the iteration, PWE assigns overwhelming traffic flow on several high-capacity edges, causing edge latencies to be higher than those of SO. 
It can be inferred that the congestion is likely to occur on those edges with the overwhelming flow. 
Fig. \ref{edgeutilization} shows that those edges with significantly high traffic latencies are severely overloaded, which indicates that the evacuation process is highly disrupted.

\section{Conclusion}

In this paper, we have formulated a Stackelberg game framework to quantify and analyze the impact of informational attacks that aim to manipulate the traffic data to mislead the Online Navigation Platforms (ONP) to provide users with falsified route recommendations.

Through sensitivity analysis, we have shown the continuity and differentiability of the attack utility function and characterized its smoothness level. 
The result has shown that the PPoA is a $C^1$-function with respect to the poisoning attack parameters, and an optimal strategy of the attack model can be achieved by a consistent Stackelberg learning process. 
It reveals the vulnerabilities of the Wardrop Equilibrium (WE)-based flow planning systems and showcases the disruptive effects that an attacker can inflict on the entire traffic network. 

Future research directions would include the investigation of the poisoning of transient equilibrium formation behavior and the development of  effective defensive and detective strategies against this class of attacks.

\bibliography{ref}
\bibliographystyle{splncs04}

\appendix

\section{Path Flow IFT} \label{appendixa}

In the path flow poisoning scenario, the parameterized Lagrangian can be written as:
\begin{equation*}
    L(\mu, \lambda, \nu,  \theta, d) = \sum_{e \in \mathcal{E}} \int_{0}^{(\Delta \Phi_{\theta} \mu)_e}  \ell_e (z)dz - \lambda^{\top} \Phi_{\theta} \mu  + \nu^{\top} ( \Lambda \Phi_{\theta} \mu - \Phi_dQ) .
\end{equation*}
Similarly for a candidate solution $\tilde{\mu}, \tilde{\lambda}, \tilde{\nu}$, write down the KKT conditions as:
\begin{equation*}
    \begin{aligned}
          - \Phi_{\theta} \mu & \preceq 0 \\
           \Delta \Phi_{\theta} \mu - \Phi_dQ & = 0 \\
           \tilde{\lambda}_p & \geq 0, \quad p = 1, \ldots, | \mathcal{P}| \\
           \tilde{\lambda}_p (\Phi_{\theta} \mu)_p & = 0, \quad p = 1, \ldots, | \mathcal{P}| \\
           \sum_{e \in \mathcal{E}} (\Delta\Phi_{\theta})^{\top}_{pe} \ell_e((\Delta \Phi_{\theta} \mu)_e) - (\Phi_{\theta}^{\top} \lambda)_p + (\Phi_{\theta}^{\top} \Lambda^{\top} \nu)_p & = 0 , \quad p = 1, \ldots, | \mathcal{P}|
    \end{aligned}
\end{equation*}

Again, we define the parameterized function $g(\tilde{\mu}, \tilde{\lambda}, \tilde{\nu}, \theta, d )$ 
\begin{equation}
    \label{gfuncforpath}
    g(\tilde{\mu}, \tilde{\lambda}, \tilde{\nu}, \theta, d ) = 
    \begin{bmatrix}
    \nabla_{\mu} L(\mu, \lambda, \nu, \theta,d) \\
    - \operatorname{diag}(\lambda) \Phi_{\theta} \mu \\
    \Delta \Phi_{\theta} \mu - \Phi_d Q
    \end{bmatrix}
\end{equation}

The partial Jacobian of variable $(\tilde{\mu}, \tilde{\lambda}, \tilde{\nu})$ is 
\begin{equation*}
    \mathrm{D}_{(\tilde{\mu}, \tilde{\lambda}, \tilde{\nu})} g(\tilde{\mu}, \tilde{\lambda}, \tilde{\nu}, \theta, d)=
    \begin{bmatrix}
    \mathrm{D}_{\mu} \nabla_{\mu} L(\tilde{\mu}, \tilde{\lambda}, \tilde{\nu}, \theta, d) &  \Phi_{\theta}^{\top} & (\Delta \Phi_{\theta})^{\top} \\
    \operatorname{diag}(\tilde{\lambda}) \Phi_{\theta} & \operatorname{diag}(- \Phi_{\theta}\mu) & 0 \\ 
    \Delta \Phi_{\theta}  & 0 & 0\end{bmatrix}
\end{equation*}

And the partial Jacobian for $\theta$ and $d$ is 

\begin{equation}
     \mathrm{D}_{(\theta, d)} g(\tilde{\mu}, \tilde{\lambda}, \tilde{\nu}, \theta, d)=
    \begin{bmatrix}
    \mathrm{D}_{\mu} \nabla_{\mu} L(\tilde{\mu}, \tilde{\lambda}, \tilde{\nu}, \theta, d) &  \Phi_{\theta}^{\top} & (\Delta \Phi_{\theta})^{\top} \\
    \operatorname{diag}(\tilde{\lambda}) \Phi_{\theta} & \operatorname{diag}(- \Phi_{\theta}\mu) & 0 \\ 
    \Delta \Phi_{\theta}  & 0 & 0
    \end{bmatrix} .
\end{equation}

We omit the explicit gradient calculation as there are diverse possibilities of parameterization.
Note that in this formulation, the conditions for $\mathrm{D}_{(\tilde{\mu}, \tilde{\lambda}, \tilde{\nu})}g$ to be non-singular becomes $ \operatorname{diag}(-\Phi_{\theta} \mu)$ and $(\Delta \Phi_{\theta})$ being invertible. A result like Theorem \ref{iftbeckman} can be derived using a similar analysis.

\section{Sketch of Proofs for Sensitivity Analysis} \label{appendixb}
We omit the proofs of Lemma \ref{charfeasibility} and \ref{continuity} (a), which is adapted from Lemma 8.3 and Corollary 8.1 of \cite{still2018lectures} by inserting $\Phi_{\theta}$ and $\Phi_d$. The following proofs are based on these preliminary results.
\begin{proof}[Lemma \ref{continuity} (b)]
  It suffices to show the smoothness of $\langle q^*(\theta, d) \ell (q^* (\theta, d)) \rangle$, for $(\theta_1, d_1), (\theta_2, d_2) \in \Theta \times \mathcal{D}$, denote variable $z_1 =(\theta_1, d_1)$ and $z_2 =(\theta_2, d_2)$, by triangular inequality and Cauchy-Schwarz inequality,  
  \begin{equation*}
      \begin{aligned}
         & \| \langle q^*(z_1 ) \ell (q^* (z_1)) \rangle -   \langle q^*(z_2) \ell (q^* (z_2)) \rangle  \|  \\  \leq  & \| \langle q^*(z_1 ) \ell (q^* (z_1)) \rangle -   \langle q^*(z_1) \ell (q^* (z_2)) \rangle  \| + \| \langle q^*(z_1 ) \ell (q^* (z_2)) \rangle -   \langle q^*(z_2) \ell (q^* (z_2)) \rangle  \| \\
         \leq & \|q^*(z_1)  \| \| \ell(q^* (z_1)) - \ell(q^* (z_2)) \| + \|\ell (q^* (z_2))\|\|q^*(z_1) - q^*(z_2) \| \\
         \leq & \sqrt{|\mathcal{E}|}D l_q l_0 \|z_1 - z_2 \| + \sqrt{|\mathcal{E}|} l_q c_0  \| z_1 - z_2\| .
      \end{aligned}
  \end{equation*}
  \qed
\end{proof}

\begin{proof}[Lemma \ref{iftbeckman}]
  Immediately follows substituting the condition $H(x, t) = 0$ in general IFT with Stationarity KKT condition $g(q, \lambda, \theta, d) = 0$, and checking the Shur complement of partial Jacobian $\mathrm{D}_{(q, \lambda)} g(\tilde{q}, \tilde{\lambda}, \theta, d)$.
  \qed
\end{proof}

\begin{proof}[Theorem \ref{attackgrad}]
  It suffices to show for variable $\theta$. Taking derivative gives:
  $$
  \nabla_{\theta} \mathcal{L} = \theta - vec(I_{|\mathcal{E}|}) - \frac{\gamma}{S^{\star} } \langle \nabla_{\theta}q^*,  \ell (q^* )\rangle + \langle  q^* , \nabla_{\theta} q^* \mathrm{D} \ell(q^*)\rangle
  $$
  Rearranging the terms yields the results. 
  \qed
\end{proof}

\begin{proof}[Theorem \ref{attackobjlip}]
For the attack cost term, we can compute the Lipschitz constant with respect to the two variables as $ \sqrt{| \mathcal{E}|} \|\theta_1 - \theta_2 \|$ and $\sqrt{| \mathcal{W}|} \|d_1 - d_2 \|$, respectively. Thus the first part for the constant should be $\sqrt{2}\max\{\sqrt{| \mathcal{E}|},  \sqrt{| \mathcal{W}|} \} = \sqrt{2}\sqrt{| \mathcal{E}|}$. For the second part, multiplying the constant in Lemma \ref{continuity} (b) with $S^{\star}$ and $\gamma$ yields the result.
  \qed
\end{proof}

\begin{proof}[Lemma \ref{gradientlip}]
We proceed under the assumption of boundedness and Lipschitz smoothness of $\|\nabla_{\theta} q^*\|_{op}$. We analyze two terms, $ \langle \nabla_{\theta}q^*,  \ell (q^* )\rangle$ and $\langle  q^* , \nabla_{\theta} q^* \mathrm{D} \ell(q^*)\rangle$. Write $q^*(\theta_1, d)$ and $q^*(\theta_2, d)$ as $q^*_1$ and $q^*_2$, respectively. 
For the first term, we have 
\begin{align*}
   &  \|\langle  \nabla_{\theta}q^*_1,  \ell (q^*_1 )\rangle - \nabla_{\theta}q^*_2,  \ell (q^*_2 )\rangle\| \\
\leq    &   \|  \langle\nabla_{\theta}q^*_1,  \ell (q^*_1 )\rangle - \nabla_{\theta}q^*_1,  \ell (q^*_2 )\rangle\| +  \| \langle \nabla_{\theta}q^*_1,  \ell (q^*_2 )\rangle - \nabla_{\theta}q^*_2,  \ell (q^*_2 )\rangle\|  \\
\leq & (C_0 l_0 l_q  \sqrt{|\mathcal{E}|} + C_1 c_0 \sqrt{|\mathcal{E}|} ) \| \theta_1 - \theta_2\|.
 \end{align*}
 For the second term, by the monotonicity of $\ell$ and the sensitivity results,
 \begin{align*}
     & \| \langle  q^*_1 , \nabla_{\theta} q^*_1 \mathrm{D} \ell(q^*_1)\rangle - \langle  q^*_2 , \nabla_{\theta} q^*_2 \mathrm{D} \ell(q^*_2)\rangle \| \\
     \leq & \| \langle  q^*_1 , \nabla_{\theta} q^*_1 \mathrm{D} \ell(q^*_1)\rangle - \langle  q^*_1 , \nabla_{\theta} q^*_2 \mathrm{D} \ell(q^*_2)\rangle\| + \|\langle  q^*_1 , \nabla_{\theta} q^*_2 \mathrm{D} \ell(q^*_2)\rangle - \langle  q^*_2 , \nabla_{\theta} q^*_2 \mathrm{D} \ell(q^*_2)\rangle\| \\
     \leq & \left(\sqrt{|\mathcal{E}|} D (C_0\sqrt{|\mathcal{E}|}l_1 l_q + C_1 \ell^{\prime}(D) \sqrt{|\mathcal{E}|}  ) + l_q C_0 \sqrt{|\mathcal{E}|} \ell^{\prime}(D) \right) \| \theta_1 - \theta_2\|
 \end{align*}
 Summing the two terms together yields the result.
  \qed
\end{proof}

\section{Sketch Proof of Proposition \ref{gradefficiency}}\label{propproof}
\begin{proof}
We show the sample bound for $\nabla_{\theta}\mathcal{L}$ approximation, the proof of sample bound for $\nabla_d\mathcal{L}$ follows the similar procedure.
  Let $\hat{\nabla}$ denote the sample average in \eqref{sampleavg}, the approximation error can be broken into two terms:
  \begin{equation*}
  \begin{aligned}
         \hat{\nabla}-\nabla_{\theta} \mathcal{L}((\theta, d), q^*)=\nabla_{\theta} \mathcal{L}^{\theta}_r((\theta, d), q^*)-\nabla_{\theta} \mathcal{L}((\theta, d), q^*) +  \hat{\nabla}-\nabla_{\theta}\mathcal{L}^{\theta}_r((\theta, d), q^*) 
  \end{aligned}
  \end{equation*}
  For the first term, choose $h_{r}(1 / \epsilon)=\min \left\{1 / r_{0}, 2 L_1  / \epsilon\right\}$, by Lemma \ref{gradientlip} when $r < 1/h_r(1/\epsilon) = \epsilon / 2 L_1$, $\| \nabla_{\theta} \mathcal{L} ((\theta + u, d), q^{*}) - \nabla_{\theta}\mathcal{L} ((\theta, d), q^{*})\| \leq \epsilon /4$. Since $$\nabla_{\theta} \mathcal{L}_r^{\theta} ((\theta, d), q^*) =  \nabla_{\theta}  \mathcal{L}^{\theta}_r((\theta,d), q^* ) = \frac{dim(\Theta)}{r^2}\E_{ u \sim  \mathbb{S}^{\theta}_r} [\mathcal{L}((\theta + u, d), q^*) u],$$
  by triangular inequality, $\| \nabla_{\theta} \mathcal{L}^{\theta}_r ((\theta, d), q^{*}) - \nabla_{\theta} \mathcal{L} ((\theta, d), q^{*})\| \leq \epsilon /2$. 
  
  Select $r_0$ such that for any $u \sim \mathbb{S}_r$, it holds that $ \mathcal{L}((\theta + u, d), q^*)$. By Theorem \ref{attackobjlip}, one can select such a $r_0$ by examining related constants. 
   Since $ \E[\hat{\nabla}] = \nabla_{\theta}  \mathcal{L}^{\theta}_r((\theta,d), q^* )$, and each sampled norm is bounded by $ 2 dim(\Theta) \mathcal{L} /r$,
   by vector Bernstein's inequality, when $ m \geq h_{sample}(d, 1/\epsilon) \propto d (\frac{d \mathcal{L}^2}{\epsilon r}) \log d / \epsilon$, with probability at least $ 1 - (d/\epsilon)^{-d}$, we have 
   \begin{equation*}
       \| \hat{\nabla} - \nabla_{\theta}\mathcal{L}^{\theta}_r((\theta, d), q^*)\| \leq \epsilon/2,
   \end{equation*}
    hence proving the claim.
    \qed 
\end{proof}

\end{document}